\documentclass[aps,prb,twocolumn,superscriptaddress,amsfonts,amsmath,amssymb]{revtex4}
\usepackage{bm}
\usepackage{graphicx}
\usepackage{amsmath}
\usepackage{amstext}
\usepackage{amssymb}
\usepackage{amsfonts}
\usepackage{amsbsy}

\newcommand{\Eq}[1]{Eq.~(\ref{#1})}

\begin{document}

\title{On the stability of $U(1)$ spin liquids in two dimensions}
\author{Michael Hermele}
\affiliation{Department of Physics, University of California, Santa 
Barbara, California 93106}
\author{T. Senthil}
\affiliation{Department of Physics, Massachusetts Institute of 
Technology, Cambridge, Massachusetts 02139}
\author{Matthew P. A. Fisher}
\affiliation{Kavli Institute for Theoretical Physics, University of 
California, Santa Barbara, California 93106}
\author{Patrick A. Lee}
\affiliation{Department of Physics, Massachusetts Institute of 
Technology, Cambridge, Massachusetts 02139}
\author{Naoto Nagaosa}
\affiliation{CREST, Department of Applied Physics, University of Tokyo, 
Tokyo 113, Japan}
\author{Xiao-Gang Wen}
\affiliation{Department of Physics, Massachusetts Institute of 
Technology, Cambridge, Massachusetts 02139}
\date{\today}
\begin{abstract}
We establish that spin liquids described in terms of 
gapless fermionic (Dirac) spinons and gapless $U(1)$ gauge
fluctuations can be stable in two dimensions, at least when the physical $SU(2)$ spin symmetry is
generalized to $SU(N)$.  Equivalently, we show that compact QED$_3$ has a deconfined phase for a large number of fermion fields, in the sense that monopole fluctuations can be irrelevant at low energies.
A precise characterization is provided by an emergent global topological $U(1)$ symmetry corresponding
to the conservation of gauge flux.
 Beginning with an $SU(N)$ generalization of the $S=1/2$ square lattice Heisenberg antiferromagnet, we consider the $\pi$-flux spin liquid and, via a systematic analysis of all operators, show that there are
\emph{no} relevant perturbations (in the renormalization group sense) about the large-$N$ spin liquid fixed point, which is thus a stable phase.  We provide a further illustration of this conclusion with an approximate renormalization group calculation that treats the gapless fermions and the monopoles on an equal footing.  This approach directly points out some of the flaws in the erroneous ``screening" argument for the relevance of monopoles in compact QED$_3$.
\end{abstract}
\maketitle

\section{Introduction}
\label{sec:intro}

Phases of strongly correlated electrons in two dimensions exhibit a remarkable
array of unusual and interesting behavior.  Quantum spin liquids\cite{pwa}
(Mott insulators with no broken symmetries) are a particularly exotic class of
states that may play a critical role in our understanding of several
interesting materials.  While much theoretical progress has been made in
understanding the universal properties of such states, many basic questions
remain unanswered.  One of these concerns the stability of spin liquids
described at low energies by gapless fermionic $S=1/2$ spinons and a
fluctuating $U(1)$ gauge field.  We focus here on the cases where the spinons
have a linear dispersion in the vicinity of discrete Fermi points and can be
described as Dirac fermions.  One example of such a phase, the staggered flux
(sF) spin liquid proposed in Ref.~\onlinecite{affleck-marston-rapid}, may play a key
role in understanding the underdoped cuprate superconductors 
\cite{WLsu2,KL9930,RWspin,FT0103}.

In this class of $U(1)$ spin liquids the fermions and gauge fluctuations 
interact strongly at low energies, and there are no well-defined low-energy 
quasiparticles\cite{appelquist,appelquist-heinz,RWspin}$^,$
\footnote{Strictly speaking, the sF spin liquid is not characterized by
the presence of gapless fermionic spinons and gapless gauge bosons, since they
are not well defined.  Instead the sF spin liquid is characterized by the
algebraic correlations between various spin operators.  The sF spin liquid is
also called algebraic spin liquid to stress this point \cite{RWspin}.}.
Therefore the issue of whether these states can be stable is in general
difficult to resolve, but becomes tractable, for example, when the physical
$SU(2)$ spin is generalized to $SU(N)$ and $N$ is taken 
large\cite{affleck-marston-rapid}. The problem can be formulated in terms 
of $N$ species of Dirac fermions coupled to a compact $U(1)$ gauge field in two 
spatial dimensions, often referred to as compact QED$_3$. 
The gauge fluctuations are suppressed at large $N$ and calculation within 
the framework of a $1/N$ expansion is possible, but even in this case there has been 
significant controversy. Based on an analysis of the monopole 
fluctuations\cite{ioffe-larkin} and the symmetries in
the sF state \cite{xgw-qorder-ssl}, it was argued that the sF spin liquid is a
stable phase and the gapless spin excitations are protected \cite{RWspin} at
least in the large $N$ limit.  However, Refs.~\onlinecite{sachdev-park,herbut-seradjeh,HSS0310} 
argue that, due
to a screening effect in 3D, the
monopole fluctuations always result in the confinement of the $U(1)$ gauge
field, thus destabilizing the sF state and its close relatives.  In this 
paper we resolve the ongoing debate and argue rather rigorously that
the compact QED$_3$ problem is deconfined for large $N$ and that
at least some $U(1)$ spin liquids are indeed stable in two dimensions.  Whether similar 
deconfinement also obtains in physically important $SU(2)$ spin models is left open for future work.

These spin liquids generally possess at low energies
much higher symmetry than the microscopic spin models in which they arise.  Of
particular importance is an emergent topological global $U(1)$ that is
associated in the gauge theory description with conservation of the gauge
flux. In a \emph{compact} $U(1)$ gauge theory flux is only conserved modulo $2\pi$ due
to the presence of monopole fluctuations, so this emergent symmetry is equivalent to the
irrelevance of monopoles at low energy. When monopoles are \emph{relevant} it is believed that
confinement is inevitable, so this symmetry provides precise meaning to the notion of
deconfinement of spinons in the spin liquid. We emphasize that our 
use of the term `deconfinement' does not
imply that the spinons are to be thought of as free at low energies.
Roughly speaking, they are instead as free as possible given their strong interaction with
the gauge fluctuations. A close parallel exists with the deconfined quantum critical
points\cite{dqcp-science,dqcp-longpaper} that have been studied recently where
an emergent topological global $U(1)$ also characterizes the low energy fixed
point. Indeed the spin liquids discussed in the present paper may be viewed as 
deconfined quantum critical {\em phases}. 

We shall consider an $SU(N)$ generalization of the $S=1/2$ Heisenberg model on
the square
lattice\cite{affleck-marston-rapid,affleck-marston-prb,arovas-auerbach}.  We
define the model in terms of slave fermions $f_{{\bf r}\alpha}$ ($\alpha =
1,\dots,N$) transforming in the fundamental representation of $SU(N)$.
Choosing $N$ even and imposing the local constraint  
$f^\dagger_\alpha f^{\vphantom\dagger}_\alpha = N/2$ puts the spin in the
antisymmetric self-conjugate representation.  Defining $S_{\alpha\beta} =
f^\dagger_\alpha f^{\vphantom\dagger}_\beta - \frac{1}{N} \delta_{\alpha\beta}
f^\dagger_\gamma f^{\vphantom\dagger}_\gamma$, the Hamiltonian is
\begin{equation}
\label{spin-hamiltonian}
{\cal H}_{SU(N)} = \frac{J}{N} \sum_{\langle{\bf r}{\bf r}'\rangle}
	S_{\alpha\beta}({\bf r})\, S_{\beta\alpha}({\bf r}') \text{,}
\end{equation}
where the sum is over pairs of nearest-neighbor sites and the exchange interaction is assumed positive,
$J>0$.  For $N=2$ this reduces to the familiar
$S=1/2$ Heisenberg antiferromagnet, with global $SU(2)$ spin rotation
symmetry.

This model can be solved in the $N \to \infty$ limit, where slave fermion
mean-field theory becomes exact.  The corresponding mean field Hamiltonian is
quadratic and describes the hopping of  fermionic spinons with a hopping
matrix element that is determined self-consistently. It is known that among
mean field solutions that preserve all lattice symmetries the energetically
favored one has a flux of $\pi$ through every plaquette of the square lattice
that is seen by the spinons. This state, known as the $\pi$-flux ($\pi$F)
state, may also be made the global minimum energy mean field state upon
addition of suitable biquadratic\cite{affleck-marston-prb} or ring exchange\cite{xgw-origin-light}
terms. Here we are only concerned with \emph{stability} of the $\pi$F state, which is a
universal feature independent of such microscopic parameters.
At the mean field
level the spinon dispersion has two distinct gapless Fermi points. 
Linearizing the dispersion in the vicinity of these
points provides a description of the low energy physics in terms of a
continuum theory of $2N$ species of $2$-component Dirac fermions.

The crucial question is the fate of this picture upon including fluctuations
beyond the mean field.  In the large-$N$ spin model the primary effect of
fluctuations is to induce a coupling of the fermionic spinons to a {\em
compact} $U(1)$ gauge field.    In $(2+1)$ dimensions the compactness means
that there are pointlike instantons (also known as monopoles) in space-time.
At any such monopole event the total gauge flux associated with the $U(1)$
gauge field changes by an integer multiple of $2\pi$.  In pure $U(1)$ gauge
theories ({\em i.e.} without any dynamical spinon fields) such instantons
always proliferate and lead to confinement\cite{Polyakov}.  This then raises
the question of whether the spinons of the mean field $\pi$F state can
escape confinement once the coupling to the compact $U(1)$ gauge field is
included. In contrast to some recent 
studies,\cite{sachdev-park,herbut-seradjeh,HSS0310}
we will argue that in the large $N$ limit spinon
deconfinement will indeed obtain.

Consider first a description of the low-energy physics of the $\pi$F state
that {\em ignores the compactness of the $U(1)$ gauge field}. The appropriate
effective field theory contains $2N$ flavors of two-component Dirac fermions
$\psi_j$ (j=1,2,...$,2N$) minimally coupled to a \emph{noncompact} $U(1)$ gauge field $a_\mu$
(with $\mu = x,y,\tau$).  After rescaling spatial coordinates to set the Dirac fermion velocity to unity, the
imaginary time action $S=\int d\tau d^2 x {\cal L}$ is given simply by
\begin{equation}
\label{et}
{\cal L} = \bar{\psi}_j \gamma^\mu (\partial_\mu +i a_\mu) \psi_j
+ \frac{1}{8\pi e^2} (\epsilon_{\mu\nu\lambda}\partial_\nu a_\lambda)^2 \text{,}
\end{equation}
where the $\gamma^\mu$ are $2 \times 2$ matrices satisfying
$\{\gamma^\mu, \gamma^\nu \} = 2\delta^{\mu \nu}$.
We have included an explicit Maxwell term for the gauge field, which will be
generated by integrating out short wavelength modes. Earlier work has argued
that this theory is critical and flows to a scale-invariant fixed point for
$N$ sufficiently large. The question then is whether properly accounting for
the compactness of the gauge field, and including all other perturbations allowed 
by the microscopic symmetries, destabilizes this scale-invariant fixed point --
in other words are monopoles (or some other allowed operator) 
a relevant perturbation in the renormalization group sense?

The issue of stability to monopole fluctuations is subtle, and before attacking the problem with
fermions it
will be useful to briefly recap Polyakov's argument for confinement in
\emph{pure} compact $U(1)$ gauge theory. By analogy with ordinary electrostatics, it is clear that
monopoles will have a $1/r$ interaction in space-time coming from the Maxwell term in 
the above Lagrangian.  If we let integer $Q_i$ represent the monopole number sitting on the sites
of a 3D (say cubic) lattice, the effective action for the monopole gas will be given by,
\begin{equation}
S_m = \frac{1}{2} \sum_{i \ne j} Q_i Q_j V(\vec{r}_i - \vec{r}_j) + s_c \sum_i Q_i^2 ,
\label{Smonopole}
\end{equation}
with $V(r) = e_m^2/r$ (for large $r$),  where we have defined a ``magnetic charge,"
$e_m = 1/2e$.  In the second term above, $s_c$ represents the action cost for the monopole cores, which will depend on short distance physics.  Since a single monopole costs a finite action, due to entropic effects
one expects them to always proliferate and be in a ``plasma" phase.  This can be simply established by
decoupling the monopole interaction term using a Hubbard-Stratanovich field $\chi_i$, and then tracing out the monopoles on each site.  For $E_c >>1$ one thereby obtains,
\begin{equation}
S_{sg} = \frac{1}{2} \int \frac{d^3q}{(2\pi)^3} V^{-1}(q) |\chi(q)|^2 - z \sum_i \cos(\chi_i) ,
\end{equation}
where $V(q)= 4\pi e_m^2/q^2$ and $\chi(q)$ are Fourier transforms of $V(r)$ and $\chi_i$,
respectively, and $z =2e^{-s_c}$.  Back in real space
the sine-Gordon action can be written in terms of a Lagrangian density, $S_{sg} = \int d^3 r {\cal L}_{sg}$, which in the continuum limit is simply
\begin{equation}
{\cal L}_{sg} = \frac{1}{8\pi g} (\partial_\mu \chi)^2 - z \cos(\chi)   ,
\end{equation}
where we have defined a coupling constant $g = e_m^2$.  In a Hamiltonian picture,
the operator $e^{i \chi}$ adds $2\pi$ magnetic flux, and can be thought of as a magnetic-flux
creation operator.  In space-time it creates a monopole, so that the coupling
$z$ is correctly interpreted as the monopole fugacity.  Implementing a simple momentum-space
renormalization group (RG)  perturbatively up to second order in $z$ gives the RG flow equations, 
\begin{equation}
\frac{\partial z}{\partial \ell}  = (3- \frac{g \Lambda}{\pi} ) z
\label{RG-z}
\end{equation}
\begin{equation}
\frac{\partial  g}{\partial \ell}  = -g - \frac{c g^3 z^2}{\Lambda^4}\text{,}
\label{RG-g-mon}
\end{equation}
with $\ell$ the usual logarithmic length rescaling and $\Lambda$ a high momentum ultraviolet cutoff.
Here $c$ is a cutoff-dependent positive dimensionless constant.
The combination $\Delta_m = g\Lambda/\pi$ is the effective scaling dimension of the monopole creation operator
$e^{i\chi}$, and relevance/irrelevance of $z$ depends as usual on whether
the scaling dimension is larger/smaller than the space-time dimenson $D=3$.
Since the coupling $g$ scales to zero, the monopole fugacity clearly grows,
indicating a proliferation of monopoles on long length scales.  The field $\chi$ gets pinned
in the minimum of the cosine potential (a ``smooth" phase), which dominates over the gradient term and   generates a mass for $\chi$.  This corresponds to a spatially short-ranged effective interaction
potential, so the monopole gas is clearly in a plasma phase
that can screen effectively.  Due to the proliferation of ``magnetically" charged monopoles,
this corresponds to the phase of the $U(1)$ gauge theory which confines the ``electric charges" --
in this case the fermionic spinons.

We now turn to the effects of the {\it gapless} fermions on the issue of monopole proliferation.
Existing arguments for and against stability of the monopole-free critical theory begin with a kind of
random-phase approximation (RPA), where one integrates out the fermions to obtain an effective
action for the gauge field. Denoting by $a_t$ the transverse part of the gauge field, this
action has the following highly schematic form:
\begin{equation}
\label{rpa-action}
S_{\rm eff}[a_\mu] = N \Big(
\int d^3 q |{\bf q}| |a_t|^2 +
\int (d^3 q)^3 \frac{1}{|{\bf q}|} |a_t|^4 + {\cal O}(a_t^6) \Big)
\end{equation}
As $N \to \infty$ one has $a_t \sim 1/\sqrt{N}$ for fluctuations with finite action,
and the nonlinear terms are \emph{apparently} higher-order in $1/N$
than the leading Gaussian term; the action is therefore truncated at Gaussian order.
For this quadratic theory, the effective interaction between monopoles
is very long-ranged, $V_{\rm RPA}(r) \sim \ln(r)$, {\it growing} logarithmically in the space-time separation.  Moreover, the action of a single monopole configuration diverges logarithmically in the
infrared (the system size, say).  The coefficient of the logarithm is proportional to $N$ 
and very na\"{\i}vely can be equated with 
the scaling dimension $\Delta_m$
of the monopole creation operator (however, see later).  The argument for the 
stability of the $\pi$F phase is then
based on two observations. (a) Since the (apparent)
scaling dimension of the monopole creation operator in the large $N$ limit is (much) greater than
the space-time dimension, monopole  fluctuations are irrelevant\cite{ioffe-larkin,kns-prl},
and a na\"{\i}ve 
perturbation theory that includes the monopoles should be free
of infrared divergences. (b) The global lattice and spin rotation symmetries of the underlying  
spin model impose
constraints on the continuum (non-compact) theory of the $\pi$F
state, precluding, for example, a mass term for the fermions. This is a \emph{projective symmetry} described by a projective
symmetry group \cite{xgw-qorder-ssl}. If this symmetry is unbroken, it protects the
gaplessness of the spinons and the $U(1)$ gauge bosons against all
perturbative fluctuations
within the monopole-free sector of the theory\cite{xgw-qorder-ssl,WZqoind}.

More recent work has, however, argued that this reasoning is too na\"{\i}ve 
and raised important and troubling questions.  Indeed,
beginning with the same quadratic gauge field action obtained within RPA by integrating out the
fermions, it has been argued that the ``bare" interaction between monopoles,
$V_{\rm RPA}(r) \sim \ln(r)$, 
will be screened by fluctuating monopole-antimonopole pairs present at finite monopole density.
Specifically, screening was argued to reduce the logarithmic interaction to a $1/r$ form
-- the same potential that arises from the Maxwell term alone in the absence of any 
fermions\cite{sachdev-park,herbut-seradjeh}.  Then Polyakov's argument (shown above) was invoked
to conclude that
monopoles would always proliferate and lead to spinon confinement.

This argument was explicitly demonstrated in the recent RG treatment of Herbut and
Seradjeh\cite{herbut-seradjeh}, who effectively treated a gas of monopoles
interacting via a pairwise logarithmic interaction, as in Eq.~(\ref{Smonopole})
except with $V(r) \rightarrow V_{\rm RPA}(r)$.  They made explicit use of the 
duality transformation to arrive at a sine-Gordon theory as above, but with an anomalous
kernel.  They argued that the presence of the monopole fugacity term will lead to a self-energy
correction at second order, $V_{\rm eff}^{-1}(q) \rightarrow V_{\rm RPA}^{-1}(q) + \Sigma(q)$,
with a self-energy $\Sigma(q) \sim z^2 q^2$ at small $q$.
The effective ``screened" interaction between monopole pairs is then simply
\begin{equation}
V_{\rm eff} (q) = \frac{V_{\rm RPA}(q)}{1 + \Sigma(q) V_{\rm RPA}(q)},
\end{equation}
which takes 
an intuitive 
``RPA-like" form with $\Sigma(q)$ the monopole
density-density correlation function.
Since $V_{\rm RPA}(q) \sim q^{-3}$,
the $\Sigma(q) V_{\rm RPA}(q)$ term dominates in the denominator, giving 
$V_{\rm eff}(q) \sim 1/\Sigma(q) \propto 1/q^2$ or $V_{\rm eff}(r) \sim 1/r$.
Following Polyakov's original argument \cite{Polyakov}, the $1/r$ interaction is further screened
to become short ranged,
signaling a proliferation of monopole events and confinement. It appears that 
regardless of the value of $N$,
the proliferation of monopoles destabilizes the $\pi$F state and leads to spinon confinement.

A first hint at the fallacy of this argument is that in the presence of monopoles the usual RPA
treatment of the gauge interaction obtained by integrating out the fermions to {\it quadratic} order in $a_\mu$ is not correct,
even in the large-$N$ limit\footnote{This has been discussed in
Ref.~\onlinecite{dqcp-longpaper} for a class of models with bosonic matter.
It is trivial to extend our argument to these cases, where it provides further
support for the existence of generic deconfined quantum critical points.
Ref.~\onlinecite{dqcp-longpaper} also anticipated that the screening
argument is not to be trusted in fermionic
models such as that considered here, and speculated that as a result
some $U(1)$ spin liquids might exist as stable phases in two dimensions.}.
This is because while the gauge
fluctuations are indeed suppressed at large-$N$, they are suppressed
\emph{with respect to some classical background configuration}.  Monopoles are
drastic perturbations to the uniform $a_\mu = 0$ background, so in principle
one needs to do a different fermion functional integral for every configuration
of monopoles. Another way to say this is that, in the effective action Eq.~(\ref{rpa-action}),
the finite-action $N \to \infty$ fluctuations about a static background of monopoles are not
controlled only by the Gaussian term. This happens precisely because this action is written as
an expansion about the \emph{wrong} classical background.
The same situation occurs in the large-$N$ $\mathbb{CP}(N)$
model at its critical point\cite{ms} and related models\cite{dqcp-longpaper};
these are similar to the model of interest in this paper, but have bosonic
matter fields.   In all these cases, the monopoles are described by a
strongly-coupled {\it non-Gaussian} theory with a rather specific structure of
multi-monopole interactions.  There is thus no reason to believe that a
treatment based on a model as a generic Coulomb gas of monopoles with just
{\em pairwise} logarithmic interactions is legitimate.  Arguments and results
valid for such a generic gas could potentially be modified for the rather
special monopole gas that correctly obtains in the large-$N$ limit in the
present problem.

But the fallacious monopole screening argument does raise an important point.
To discuss the stability of Eq.~(\ref{et})
to monopoles it is not enough to simply show that the scaling dimension of the
monopole operator is larger than the space-time dimension. It is equally 
important to ensure that as the monopole
fugacity renormalizes towards zero, the monopoles do not induce {\it other}
operators in the monopole-free sector which are relevant. Indeed the screening occurring
in the above sine-Gordon theory ({\em i.e.} the self-energy contribution) may be viewed as
providing an explicit example where
precisely this sort of thing happens.

Thus the key to demonstrate stability of the $\pi$F state described by Eq.~(\ref{et})
is to show that there are simply no relevant perturbations with or
without the inclusion of monopoles.  In this paper, we shall give a systematic
analysis of all operators in the $\pi$F state and show that there are indeed no
relevant perturbations, at least in the large $N$ limit.

An important closely related issue is that,
in most existing discussions of this problem, scant attention has been paid
to the constraints imposed by the microscopic symmetries of the original
lattice spin model on the effective theory Eq.~(\ref{et}). Apart from global
$SU(N)$ spin rotation these symmetries include translations, rotations, parity and time-reversal, as well as ``charge conjugation," which takes the lattice fermion fields
$f^{\vphantom\dagger}_{\alpha}$ to $f^{\dagger}_{\alpha}$.  
(For $SU(2)$ spins charge conjugation is a subgroup of spin rotations,
corresponding to a rotation by $\pi$ around the $y-$axis in spin space,
but for $N>2$ is an independent discrete global symmetry of the spin Hamiltonian.)
As emphasized in Ref.~\onlinecite{xgw-qorder-ssl}, the transformations of the fermion fields and the gauge field under these symmetries define the
projective symmetry group associated with the slave fermion-gauge formulation and are crucial
in prohibiting a class of potentially relevant operators in the continuum (monopole-free)
theory.  As mentioned above, without the
projective symmetry, there is nothing to prevent the spinon mass term (a
relevant perturbation) from being induced by {\em any} generic perturbation 
to the theory (be it
inclusion of monopoles or other less drastic perturbations such as 
four-fermion interactions).

\section{General Argument}
\label{sec:general-argument}

We now present the details of our argument.  The physics of the $\pi$-flux
state and the associated fluctuations is encapsulated in the lattice gauge
theory Hamiltonian
\begin{eqnarray}
\label{lattice-gt}
&&{\cal H}_{U(1)} = \frac{h}{2} \sum_{\langle{\bf r}{\bf r}'\rangle} 
e^2_{{\bf r}{\bf r}'} \\
  &&-\,t \sum_{{\bf r} \in A} \sum_{{\bf r}' \leadsto {\bf r}}
\big[ (i + (-1)^{(r_y - r'_y)}) f^\dagger_{{\bf r}\alpha} e^{-i 
a_{{\bf r}{\bf r}'}}
  f^{\vphantom\dagger}_{{\bf r}'\alpha}
+ \text{h. c.} \big] \nonumber \\
&&+ \cdots \nonumber
\end{eqnarray}
where the ellipsis represents perturbations consistent with the symmetries.
In the second term, the first sum is over sites of the $A$ sublattice and the
second is over nearest neighbors of ${\bf r}$.  Here $a_{{\bf r}{\bf r}'}$ is
a $2\pi$-periodic vector potential living on the nearest-neighbor bonds, and
$e_{{\bf r}{\bf r}'}$ is its canonically conjugate integer-valued electric
field.  We must also specify the gauge constraint $\sum_{{\bf r}'\leadsto{\bf
r}} e_{{\bf r}{\bf r}'}  + f^\dagger_\alpha f^{\vphantom\dagger}_\alpha = 1$;
the first term is a lattice divergence of the electric field.  With this
choice Eq.~(\ref{lattice-gt}) reduces exactly to the spin model in the limit
$h/t \to \infty$ with $J \sim t^2/h$.  The apparent breaking of lattice symmetry is a gauge
artifact, which is taken care of by requiring the lattice symmetry transformations to act
on the spinons with additional gauge transformations,  thus specifying the
\emph{projective symmetry group} of this spin liquid\cite{xgw-qorder-ssl}
(see Appendix~\ref{app:symmetries}).
This information allows us to verify that the gauge theory Hamiltonian Eq.~(\ref{lattice-gt}) has the exact same global
symmetry group as the original spin Hamiltonian, and to determine what additional perturbations are
allowed.

The continuum theory is obtained by first setting $a_\mu = 0$ and solving for
the band structure of Eq.~(\ref{lattice-gt}) (this reproduces the $N = \infty$
mean field state).  Choosing a four-site unit cell, the spinon dispersion has
nodes at $(\pi/2,\pi/2)$ in the reduced Brillouin zone $k_x,k_y \in [0,\pi)$. For
each component of $SU(N)$ spin, the linear dispersion about the nodal point is described
by two 2-component massless Dirac fermions. 
Then with $N$ flavors of lattice spinons, one has a total of $2N$
flavors of 2-component Dirac fermions.  
Gauge fluctuations can
be added to this continuum theory (as in Eq.~(\ref{et})) so long as we
recognize that the gauge field is compact and allow for appropriate monopole
configurations.  It is straightforward to determine the action of the
microscopic symmetries on the continuum fields; this is outlined in Appendix~\ref{app:symmetries}.
This shows that all possible
quadratic mass terms ({\em i.e.} fermion bilinears with no derivatives) are
forbidden.  Moreover,  the velocities associated with the dispersion around
the nodes are all required to be the same.
Thus the quadratic part of the Dirac action describes $2N$ flavors of 2-component Dirac
fermions with full $SU(2N)$ symmetry. Terms that break the $SU(2N)$ symmetry
are of quartic or higher order in the fermions, and as often happens the
quadratic part has higher symmetry than the full action.

It is important to observe that if monopoles are ignored, the theory in Eq.~(\ref{et})
has in addition an extra topological global $U(1)$ symmetry, as discussed in Sec.~\ref{sec:intro}.
This
corresponds to the conservation of the gauge flux through any surface spanning
the system in space-time. A charge-$Q$ monopole event changes the flux by
$2\pi Q$ and spoils this conservation law.

Existing results in the literature show that the highly symmetric continuum
theory Eq.~(\ref{et}) flows to a conformally-invariant critical fixed point at
large-$N$. Here we examine all perturbations to this fixed point allowed in
the present problem and argue that they are irrelevant. It is convenient
to  group the perturbations into two classes: operators that do not change the
flux, and those that do. The former contains all perturbations that are
allowed in the absence of monopoles.  The latter contains the monopoles, and their
composites with polynomials in the fermion fields.

Results from the $1/N$ expansion strongly suggest that
the $SU(2N)$-symmetric noncompact theory Eq.~(\ref{et}) flows to
a critical fixed point for $N$ sufficiently large\cite{appelquist,appelquist-heinz}.
The infinite-$N$ theory is
manifestly scale-invariant, with a photon propagator proportional to $1/|q|$
at small momentum.  The effective expansion parameter for all diagrams in the
$1/N$ expansion is then dimensionless and one expects that
only logarithmic divergences will
occur; these contribute to nontrivial anomalous dimensions in principle calculable
order-by-order in $1/N$.
It should also be noted that a fermion mass cannot be generated
perturbatively in $1/N$.  (See Appendix~\ref{app:largeN} for a more detailed discussion 
on the validity of the $1/N$ expansion at large but finite $N$.)  The criticality of the theory is further supported
by a simple RG analysis in $3 - \epsilon$ dimensions\cite{senthil-unpub}, 
which finds a nontrivial
fixed point at ${\cal O}(\epsilon)$ that is presumably smoothly
connected to the large-$N$ fixed point in $d=2$. Furthermore,  
recent
large-scale numerical studies find no evidence for any symmetry breaking down to
the smallest value of $N$ simulated ($N=2$)\cite{kogut-numerics}.
This conclusion
agrees with many approximate analyses of this model,
finding $SU(2N)$ symmetry breaking only below some critical
$N_c$\cite{appelquist,appelquist-critical,nash,maris},
contradicting early claims that the symmetry is always broken\cite{pisarski,p-w-1,p-w-2}.

We next consider adding flux-preserving perturbations to Eq.~(\ref{et}) that
break its global symmetries down to those of the original lattice model. As
emphasized earlier the \emph{microscopic} symmetries forbid all possible
quadratic fermion mass terms, as well as any velocity anisotropy.  The leading
such perturbations will therefore be four-fermion terms.
At large-$N$ all $4$-fermion operators have scaling dimension $4 -
{\cal O}(1/N)$, and are hence expected to be irrelevant.
If some such operator had
instead been relevant, it could have led to spontaneous breakdown of the
$SU(2N)$ symmetry. Thus we expect that at least for $N$ large the general non-compact theory
(\emph{i.e.} with only the global symmetries of the lattice model Eq.~(\ref{lattice-gt}))
flows to an  $SU(2N)$-symmetric conformally invariant fixed point.

Finally we consider the monopole operators. Following Ref.~\onlinecite{bkw} it will
be extremely convenient to adopt a powerful point of view familiar to
conformal field theory aficionados (for a pedagogical review accessible to
condensed matter theorists see Ref.~\onlinecite{op-state}): In any $D$-dimensional
conformally invariant theory, there is a one-to-one mapping between local
operators and quantum states of the same theory quantized on the surface of
the unit sphere $S^{D-1}$.  Furthermore, energy eigenstates on the sphere
correspond to eigenoperators of RG scale transformations.  The scaling
dimension of such an operator is equal to the energy of the corresponding
quantum state. Ref.~\onlinecite{bkw} used this point of view to calculate very
simply the scaling dimension and other quantum numbers of monopole operators
for the theory in Eq.~(\ref{et}) at leading order in $1/N$.

Consider a general local operator that changes the flux by an amount $2\pi Q$.
This corresponds to a state on the sphere in a sector with a total magnetic flux of $2\pi Q$ through the surface.  The
energy of the state (and hence the scaling dimension of the corresponding
operator) is bounded below by the ground state energy in the same sector.  In the $N \to \infty$ limit the
gauge fluctuations are completely suppressed and the magnetic flux can be treated as a static, uniform background.  Furthermore, the fermions do not interact in this limit, so the problem is reduced to finding the ground state energy of $2N$ free Dirac fermions on the sphere in a background magnetic field.
As shown explicitly in Ref.~\onlinecite{bkw} (and is physically reasonable) the ground
state energy in each such sector with non-zero flux is ${\cal O}(N)$. Consequently
{\em all} flux changing operators have a scaling dimension bounded
below by a number of ${\cal O}(N)$ and are irrelevant. 
Moreover, as shown in Ref.~\onlinecite{bkw}, the scaling dimension of a charge $Q$ monopole
is {\it not} equal to $Q^2$ times the scaling dimension of the charge $Q=1$ monopole,
as is the case within the Gaussian RPA treatment in terms of
a gas of monopoles with a simple {\it pairwise} logarithmic interaction.
This makes clear that the conclusion in the monopole ``screening" argument that the logarithmic interaction
is screened down to a $1/r$ form rests on the flawed assumption of
pairwise monopole interactions.

We conclude that the effective theory in Eq.~\ref{et} flows to a conformally
invariant fixed point with no relevant perturbations  -- with or without the
inclusion of monopoles. Consequently, the $\pi$-flux $U(1)$ spin liquid state
survives as a stable deconfined gapless critical phase, at least at large-$N$.

\section{RG Analysis}

These considerations can be illustrated by the following 
(approximate) RG calculation
which further helps clarify the flaws in the RPA screening argument
in Refs.~\onlinecite{herbut-seradjeh} and~\onlinecite{HSS0310}.
We implement an approximate RG calculation following the
approach used in Ref.~\onlinecite{HSS0310}.  However,
our calculation pays attention to two important points: (a) we assume that the
fermion mass counter term is not allowed (since it is forbidden by the
projective symmetry in the $\pi$F state); (b) we let the infrared cutoff length scales for
the fermions ($L_f$) and the monopoles ($L_m$) approach 
infinity with a fixed ratio,  say $L_f/L_m=1$. This is reasonable since 
{\it both} the monopole and the fermion sectors involve long-wavelength 
and gapless degrees of freedom.  
The choice in Ref.~\onlinecite{HSS0310}, with $L_f$ taken to infinity before $L_m$, 
is not justified because it is based on an erroneous adiabatic approximation which requires
the fermions to be rapidly varying variables compared with the monopoles. 

In the absence of gapless fermions, the RG flow equations in the monopole sector
in terms of the ``running" monopole fugacity, $z(\ell)$, and the running ``magnetic charge,"
$g(\ell) = e_m^2(\ell)$, with $\ell \equiv \ln(L_m)$, are already given in Eqs.~(\ref{RG-z}) and~(\ref{RG-g-mon}).  Next consider the monopole-free sector with gapless fermions coupled to the non-compact gauge field with ``electric charge" $e$ (with Lagrangian as given in Eq.(\ref{et})).
An approximate RG perturbative in $e$ can readily be implemented on this theory, which is simply noncompact QED$_3$.  This is accomplished by simply ignoring all nonlinear terms in the action (except of course the minimal coupling vertex); in an exact treatment these would have nonzero coefficients at the fixed point of interest.  Up to order $e^6$ the RG flow equations take the simple form,
\begin{equation}
\frac{\partial(e^2)}{\partial \ell} = (4-D) e^2 -4\pi \eta \frac{N}{\Lambda} e^4 ,
\label{RG-e}
\end{equation}
with $\ell \equiv \ln(L_f)$ and $\Lambda$ an ultraviolet high momentum cutoff.
Here $\eta > 0$ is a dimensionless number.  The first term is present because the
electric charge is not dimensionless in three space-time dimensions ($D=3$), while the second term
comes from a one-loop fermion bubble screening the Coulomb interaction and reducing the effective electric charge, as is familiar from four-dimensional QED.  For large $N$ this flow equation
has a perturbatively accessible {\it stable} fixed point (with coupling $(e^*)^2= \Lambda/4\eta N$)
which describes the critical $\pi-$flux state.

At this level of approximation, where we keep only the quadratic Maxwell term in the gauge action, the
inverse relation $e_m = 1/2e$ is retained under the RG flows.  
Therefore we can deduce the effects
of the fermions on the monopoles by replacing $e^2 \rightarrow 1/4g$ in the electric charge flow Eq.~(\ref{RG-e}) above, which gives,
\begin{equation}
\frac{\partial g}{\partial \ell} = -g + \frac{\pi \eta N}{\Lambda} .
\label{RG-g-ferm}
\end{equation}
The first term represents the engineering dimension of the magnetic charge in 3D,
while the second comes from the screening of the fermions, which reduces the electric charge and thereby increases the magnetic charge. 

To arrive at a full set of RG flow equations incorporating both the monopoles and the fermions,
we must add to the right hand side of Eq.~(\ref{RG-g-ferm}) the contribution to the magnetic charge
coming from screening by monopole-antimonopole pairs, given explicitly in Eq.~(\ref{RG-g-mon}).
The full set of coupled RG flow equations for the monopole fugacity and (magnetic) charge
then take the form,
\begin{equation}
\frac{\partial z}{\partial \ell}  = (3- \frac{g \Lambda}{\pi} ) z  ,
\label{RG-z-final}
\end{equation}
\begin{equation}
\frac{\partial  g}{\partial \ell}  = -g - \frac{c g^3 z^2}{\Lambda^4} + \frac{\pi \eta N}{\Lambda} .
\label{RG-g}
\end{equation}
If $z(\ell)$ scales to zero in the $\ell \to \infty$ limit, the magnetic charge approaches a
fixed point value,
$g(\infty) \equiv  g^* = \pi \eta N/\Lambda$.
In this case the scaling dimension of the monopole creation operator becomes,
$\Delta_m = g^* \Lambda/\pi = \eta N$.  For large $N$, $\Delta_m$ is much greater than $D=3$
so that $z$ indeed does scale to zero consistent with the original assumption.
Thus, provided 
\begin{equation}
N \eta  >3, 
\label{irrcond}
\end{equation}
this approximate RG scheme predicts that the monopole fugacity scales to zero and the 
(magnetic) charge approaches a fixed point value.
The $U(1)$ gauge field is not confining at such a
fixed point, which corresponds to the $\pi$F phase with gapless spin
excitations.

One can see that the result \Eq{irrcond} is the same as that obtained
by Ioffe-Larkin \cite{ioffe-larkin} without considering the monopole
screening effect. Their analysis corresponds to the single relation in 
\Eq{RG-z-final} only, assuming $g = g^* \sim N$.
This is eventually justified in our coupled Eqs.~(\ref{RG-z-final}) and~(\ref{RG-g}).
The arguments in Refs.~\onlinecite{sachdev-park,herbut-seradjeh}
focus on the $-z^2$ term on the right side of \Eq{RG-g} which represents
screening of the magnetic charge by monopole-antimonopole pairs, a term
which scales to zero in the present treatment.

We emphasize again that the RG calculation above is approximate (even in the 
large-$N$ limit). In particular the coupling of
the gauge field to the fermions is treated in perturbation theory and 
all non-linear terms in the action are ignored. Nevertheless, it
illustrates the failure of the monopole 
screening argument in this context.

\section{Discussion}

The low energy fixed point that describes the $\pi$F spin liquid has a global
$SU(2N)$ symmetry and an extra global topological $U(1)$ symmetry. The latter
is a precise consequence of the asymptotic irrelevance of monopoles and
corresponds to conservation of gauge flux. It thus provides precise meaning to
the notion of deconfinement of the $U(1)$ gauge field.
These results may be expected to generalize to other states that also have
Dirac fermions coupled to $U(1)$ gauge fields in two dimensions such as the
staggered flux spin liquid. Our analysis, however, is controlled only at large-$N$, which simply provides a limit where these issues can be reliably addressed; future work will be required to determine whether similar deconfinement obtains in models with real $SU(2)$ spins.

We remark that in three dimensions the stability problem is essentially trivial even for $SU(2)$ spin. In this case the large-$N$ limit does not play a role; for example, it is known that $U(1)$ spin liquids can exist as stable phases in $d=3$ even if the spinons are completely gapped.  It is very likely possible to have a stable three-dimensional liquid of $SU(2)$ spins described by massless Dirac spinons interacting with an emergent $U(1)$ gauge field -- one need only show that all possible fermion mass terms are forbidden by the microscopic symmetries.  In this case the effective theory is QED in four space-time dimensions, and at low energies the spinons and photons interact only \emph{weakly}.  Because this theory is under good control and detailed predictions can be made, it may be very fruitful to search for possible realizations of such a phase in experiments and numerical simulations.

We also note that issues similar to that resolved in this paper arise in the problem of a 
Fermi surface of spinons coupled to a compact $U$(1) gauge 
field in two dimensions\cite{N9310,HSS0310}. 
Since this problem has an even
higher density of low lying excitations compared with Dirac fermions,
the
notion of integrating out all the fermions and considering monopoles 
in the resulting action truncated to quadratic order
is even more suspect.
We expect that a correct treatment will likely show that monopoles are 
irrelevant for large $N$
in this case as well.  Of course, here the monopole-free theory
is likely vulnerable to Fermi surface instabilities. It is hence unlikely to survive as the ground state 
but could perhaps be as stable as an ordinary Fermi liquid.

\begin{acknowledgments}
The authors would like to thank Leon Balents, Subir Sachdev and Ashvin Vishwanath for several helpful discussions.
This research is supported by the Department of Defense NDSEG program (M.H.),
NSF Grant No. DMR-0308945 (T.S.), NSF Grant Nos. DMR-0210790 and PHY-9907949 (M.P.A.F.),
NSF Grant No. DMR--02--01069 (P.A.L.), Grant-in-Aids and NAREGI Nanoscience Project
from the Ministry of Education, Culture, Sports, Science, and Technology (N.N.),
and NSF Grant No. DMR--01--23156 and NSF-MRSEC Grant No. DMR--02--13282 (X.G.W.).  T.S. also
acknowledges funding from the NEC Corporation, the Alfred P. Sloan Foundation,
and an award from The Research Corporation.
\end{acknowledgments}

\appendix
\section{Symmetries and Continuum Fields}
\label{app:symmetries}

For completeness, we provide here a discussion of the continuum limit of the $\pi$F mean field state and the action of the microscopic symmetries on the continuum Dirac fields.  The starting point is the mean field $\pi$F Hamiltonian
\begin{equation}
\label{eq:pi-mf}
{\cal H}_\pi = -\,t \sum_{{\bf r} \in A} \sum_{{\bf r}' \leadsto {\bf r}}
\big[ (i + (-1)^{(r_y - r'_y)}) f^\dagger_{{\bf r}\alpha} 
  f^{\vphantom\dagger}_{{\bf r}'\alpha}
+ \text{h. c.} \big] \text{.} 
\end{equation}
Note that this is simply Eq.~(\ref{lattice-gt}) with gauge fluctuations completely suppressed
($a_{{\bf r}{\bf r}'} \equiv 0$).

It is convenient to work with a four-site unit cell labeled by $({\bf R},i)$, with
${\bf R} = 2 n_x {\bf x} + 2 n_y {\bf y}$ and ${\bf r}({\bf R},i) = {\bf R} + {\bf v}_i$, where
\begin{equation} 
{\bf v}_i = \left\{ \begin{array}{ll}
{\bf 0} & i = 1 \\
{\bf x} & i = 2 \\
{\bf x}+{\bf y} & i =3 \\
{\bf y} & i = 4
\end{array} \right.
\end{equation}
The spinon operator at the site $({\bf R},i)$ is denoted $f_{{\bf R} i \alpha}$.  It is a trivial exercise to go to momentum space and solve Eq.~(\ref{eq:pi-mf}); in the reduced Brillouin zone $k_x,k_y \in [0,\pi)$
appropriate for this unit cell one finds gapless Fermi points at ${\bf Q}_0 \equiv (\pi/2,\pi/2)$.  Near this point the dispersion can be described by $2N$ 2-component Dirac fermions.  It is convenient to denote these by $\psi^A_{a \alpha}({\bf R})$.  Here $a = 1,2$ and $\alpha = 1,\dots,N$ are the $SU(2N)$ flavor indices ($\alpha$ is simply the $SU(N)$ spin index).  Also, $A =1,2$ labels the two components of each
spinor (this is often suppressed).  These fields are related to the lattice spinons as follows:
\begin{eqnarray}
\psi^1_{1\alpha}({\bf R}) &\sim&
	\frac{1}{2\sqrt{2}\ell} e^{i{\bf Q}_0\cdot{\bf R}} (f_{{\bf R}1\alpha} + f_{{\bf R}3\alpha})\\
\psi^2_{1\alpha}({\bf R}) &\sim&
	\frac{-i}{2\sqrt{2}\ell} e^{i{\bf Q}_0\cdot{\bf R}} (f_{{\bf R}2\alpha} - f_{{\bf R}4\alpha})\\
\psi^1_{2\alpha}({\bf R}) &\sim&
	\frac{-e^{-i\pi/4}}{2\sqrt{2}\ell} e^{i{\bf Q}_0\cdot{\bf R}} (f_{{\bf R}2\alpha} + f_{{\bf R}4\alpha})\\
\psi^2_{2\alpha}({\bf R}) &\sim&
	\frac{-e^{-i\pi/4}}{2\sqrt{2}\ell} e^{i{\bf Q}_0\cdot{\bf R}} (f_{{\bf R}1\alpha} - f_{{\bf R}3\alpha})\text{,}
\end{eqnarray}
where $\ell$ is the lattice spacing.

In momentum space the continuum Hamiltonian takes the form
\begin{equation}
{\cal H}_c = \int \frac{d^2 q}{(2\pi^2)} \psi^\dagger_{a\alpha}({\bf q})
\big(q_1 \tau^1 + q_2 \tau^2\big) \psi^{\vphantom\dagger}_{a \alpha}({\bf q}) \text{,}
\end{equation}
where we have chosen units to set the velocity to unity, and $\tau^i$ are the usual Pauli matrices acting
in the 2-component Dirac ``spin'' space.  Here we use the following rotated coordinates:
\begin{eqnarray}
\label{rotated-coords}
q_1 &=& \frac{1}{\sqrt{2}}(q_x + q_y) \\
q_2 &=& \frac{1}{\sqrt{2}}(-q_x + q_y) \text{.}\nonumber
\end{eqnarray}

We now simply quote the action of the microscopic symmetries on the lattice spinons and the resulting transformations for the continuum fields.  
In order to keep ${\cal H}_\pi$ invariant, in some cases the spinons transform with an additional $U(1)$ gauge transformation; this is the hallmark of projective symmetry and the projective symmetry group, and has important consequences for the action of the symmetries on the continuum fields.  We often include an extra {\em uniform} gauge transformation
$f_{{\bf r}\alpha} \to e^{i\phi} f_{{\bf r}\alpha}$ to (slightly) simplify the form of the continuum transformation laws.  It is convenient to adopt a four-component notation, defining:
\begin{equation}
\Psi_\alpha = \left( \begin{array}{c}
	\psi_{1\alpha} \\
	\psi_{2\alpha}
\end{array} \right)
\end{equation}
We can represent matrices acting in this four-component space by the tensor products of Pauli matrices
$\tau^i \mu^j$.  The $\tau^i$ matrices act in the Dirac spin space, while the $\mu^i$ act in the flavor space connecting the upper and lower 2-component spinors of $\Psi$.  In the $SU(2N)$-symmetric continuum theory, the $\mu^i$ generate the $SU(2)$ subgroup of $SU(2N)$ consisting only of emergent symmetries; these are in some sense continuous extensions of the discrete lattice symmetries.

$x$-{\em translations}. 
Translations by one lattice site in the $x$-direction act on the spinons as follows:
\begin{equation}
f_{{\bf r}\alpha} \to \exp(-\frac{i\pi}{2}\zeta_{\bf r}) f_{{\bf r}+{\bf x},\alpha}\text{,}
\end{equation}
where $\zeta_{\bf r} = 0,1,2,3$ when the coordinates of ${\bf r}$ are (even, even), (odd,even), (odd, odd) and (even, odd), respectively.
The continuum transformation law is
\begin{equation}
\Psi_\alpha \to (i\mu^1) \Psi_\alpha
\end{equation}

{\em Rotations.}  We choose to make a $\pi/2$ counterclockwise rotation about the center of a plaquette;  that is, we rotate about the
point $({\bf x}+{\bf y})/2$, which gives
${\bf r} = (r_x, r_y) \to {\bf r}' = (-r_y + 1, r_x)$.  The action on the spinon operators is:
\begin{equation}
f_{{\bf r}\alpha} \to \epsilon_{\bf r} f_{{\bf r}'\alpha}\text{,}
\end{equation}
where
\begin{equation}
\epsilon_{{\bf r}} = \left\{ \begin{array}{ll}
	+1 & {\bf r} \in A \\
	-1 & {\bf r} \in B
\end{array} \right.
\end{equation}
In the continuum we have:
\begin{equation}
\Psi_\alpha({\bf R}) \to \exp\Big(\frac{i\pi}{2}\Big(\frac{\mu_1 + \mu_2}{\sqrt{2}}\Big)\Big)
	\exp\Big(\frac{i\pi}{4}\tau^3\Big) \Psi_\alpha({\bf R}')
\end{equation}

{\em Reflections.}  We consider a
reflection of the form ${\bf r} = (r_x,r_y) \to {\bf r}' = (-r_x,r_y)$, under which the spinons transform
trivially
\begin{equation}
f_{{\bf r}\alpha} \to f_{{\bf r}'\alpha}
\end{equation}
In the continuum this leads to:
\begin{equation}
\Psi_\alpha({\bf R}) \to (i \mu^2)\exp\Big(\frac{i\pi}{2}\Big(\frac{\tau^1+\tau^2}{\sqrt{2}}\Big)\Big) 
	\Psi_\alpha({\bf R}')
\end{equation}

{\em Charge conjugation.}  As discussed in Sec.~\ref{sec:general-argument}, charge conjugation
is distinct from the $SU(N)$ spin rotation symmetry only for $N>2$.  The action on the spinons is
\begin{equation}
f_{{\bf r}\alpha} \to \exp(\frac{i\pi}{2} \zeta_{{\bf r}}) f^\dagger_{{\bf r}\alpha}
\end{equation}
In the continuum:
\begin{eqnarray}
\Psi_\alpha &\to& \Big [\Psi^\dagger_\alpha (i \tau^1)(i \mu^1)\Big]^T \\
\Psi^\dagger_\alpha &\to&  \Big[(i\tau^1)(i\mu^1) \Psi_\alpha \Big]^T
\end{eqnarray}

{\em Time Reversal.}  Time reversal is an antiunitary operation acting on the lattice 
spinons as follows:
\begin{equation}
f_{{\bf r}\alpha} \to \epsilon_{{\bf r}} f^\dagger_{{\bf r}\alpha}
\end{equation}
In the continuum this becomes:
\begin{eqnarray}
\Psi_\alpha &\to& \Big[ \Psi^\dagger_\alpha (i \tau^3) (i \mu^3) \Big]^T \\
\Psi^\dagger_\alpha &\to&  \Big[ (i \tau^3) (i \mu^3) \Psi_\alpha \Big]^T
\end{eqnarray}

It is at this point a simple exercise to show that these symmetries forbid all possible
mass terms $M_{ij} = \Psi^\dagger_{\alpha} \mu^i \tau^j \Psi^{\vphantom\dagger}_{\alpha}$,
as well as any velocity anisotropy.

\section{Validity of the $1/N$ expansion at large but finite $N$}
\label{app:largeN}

While our results are based on an analysis within the framework of the $1/N$ expansion, they should hold for large but finite $N$.  In order to see this it is important to consider the precise connection between the $1/N$ theory and actual finite-$N$ models.  

Consider the field theory Eq.~(\ref{et}) for some finite value of $N$.  We can formally integrate out the fermions to arrive at an effective action for the gauge field as in Eq.~(\ref{rpa-action}) -- one can imagine using an appropriate lattice regularization in order to ensure that this step is well-defined.  With the fermion fields gone, there is no obstacle to treating $N$ as a continuous variable and doing perturbation theory in $1/N$.  The resulting family of field theories is manifestly non-local because we have obtained it by integrating out the gapless fermions; however, for integer $N$ it is equivalent to the local field theory we started with.  For non-integer $N$ there is no reason to believe the theory is equivalent to {\em any} local theory.  The key point is that near $1/N = 0$ the points that do correspond to a local theory become arbitrarily closely spaced, so it should be a very good approximation to view the theories in a small, continuous interval near $1/N = 0$ as local.

The above considerations mean that we can think of our expansion about the $1/N = 0$ fixed point in the standard framework of the Wilsonian renormalization group, which applies only to local theories.  In this case a finite value of $1/N$ is an exactly marginal perturbation, and we can calculate corrections to the scaling dimension of any operator as $1/N$ is tuned away from zero.  The $1/N$ expansion itself gives only an asymptotic series for each of these scaling dimensions.  However, if the asymptotic $N \to \infty$ result for a given scaling dimension $\Delta$ is greater than the space-time dimension $D$, it is straightforward to use the mathematical definition of asymptotic convergence to show that $\Delta > D$ for some sufficiently large but {\em finite} value of $N$, above which the corresponding operator  is irrelevant.

\bibliography{stable-u1}

\end{document}